\documentclass[conference,10pt]{IEEEtran}
\usepackage{url}
\usepackage[utf8]{inputenc}
\usepackage{xcolor}
\usepackage{amsmath}
\usepackage{amssymb}

\usepackage[acronyms,nonumberlist,nopostdot,nomain,nogroupskip]{glossaries}
\usepackage{tablefootnote}
\usepackage{booktabs}

\usepackage{tabularx}

\usepackage{tikz}
\usepackage{pgfplots}
\pgfplotsset{compat=newest} 
\pgfplotsset{plot coordinates/math parser=false} 
\newlength\fheight
\newlength\fwidth
\usetikzlibrary{plotmarks,patterns,decorations.pathreplacing,backgrounds,calc,arrows,arrows.meta,spy,matrix}
\usepgfplotslibrary{patchplots,groupplots}
\usepackage{tikzscale}

\newif\ifexttikz
\exttikzfalse

\ifexttikz
	\usetikzlibrary{external}
\fi

\usepackage{multirow}
\usepackage{tkz-kiviat}

\usepackage[font=scriptsize]{subcaption}
\usepackage[font=footnotesize]{caption}

\usepackage{mathtools}

\newacronym{3gpp}{3GPP}{3rd Generation Partnership Project}
\newacronym{adc}{ADC}{Analog to Digital Converter}
\newacronym{5g}{5G}{5th generation}
\newacronym{aimd}{AIMD}{Additive Increase Multiplicative Decrease}
\newacronym{am}{AM}{Acknowledged Mode}
\newacronym{amc}{AMC}{Adaptive Modulation and Coding}
\newacronym{aqm}{AQM}{Active Queue Management}
\newacronym{awgn}{AGWN}{Additive White Gaussian Noise}
\newacronym{balia}{BALIA}{Balanced Link Adaptation}
\newacronym{bdp}{BDP}{Bandwidth-Delay Product}
\newacronym{bf}{BF}{Beamforming}
\newacronym{cc}{CC}{Congestion Control}
\newacronym{cdf}{CDF}{Cumulative Distribution Function}
\newacronym{cn}{CN}{Core Network}
\newacronym{cqi}{CQI}{Channel Quality Information}
\newacronym{cp}{CP}{Control Plane}
\newacronym{csirs}{CSI-RS}{Channel State Information - Reference Signal}
\newacronym{dc}{DC}{Dual Connectivity}
\newacronym{dce}{DCE}{Direct Code Execution}
\newacronym{dci}{DCI}{Downlink Control Information}
\newacronym{dl}{DL}{Downlink}
\newacronym{dmr}{DMR}{Deadline Miss Ratio}
\newacronym{dmrs}{DMRS}{DeModulation Reference Signal}
\newacronym{e2e}{E2E}{End-to-End}
\newacronym{ecn}{ECN}{Explicit Congestion Notification}
\newacronym{edf}{EDF}{Earliest Deadline First}
\newacronym{enb}{eNB}{evolved Node Base}
\newacronym{epc}{EPC}{Evolved Packet Core}
\newacronym{es}{ES}{Edge Server}
\newacronym{fdma}{FDMA}{Frequency Division Multiple Access}
\newacronym{fdd}{FDD}{Frequency Division Duplexing}
\newacronym[firstplural=Radio Access Technologies (RATs)]{rat}{RAT}{Radio Access Technology}
\newacronym{fs}{FS}{Fast Switching}
\newacronym{ftp}{FTP}{File Transfer Protocol}
\newacronym{gnb}{gNB}{Next Generation Node Base}
\newacronym{harq}{HARQ}{Hybrid Automatic Repeat reQuest}
\newacronym{hetnet}{HetNet}{Heterogeneous Network}
\newacronym{hh}{HH}{Hard Handover}
\newacronym{hol}{HOL}{Head-of-Line}
\newacronym{ia}{IA}{Initial Access}
\newacronym{imt}{IMT}{International Mobile Telecommunication}
\newacronym{iot}{IoT}{Internet of Things}
\newacronym{los}{LOS}{Line-of-Sight}
\newacronym{lte}{LTE}{Long Term Evolution}
\newacronym{m2m}{M2M}{Machine to Machine}
\newacronym{mac}{MAC}{Medium Access Control}
\newacronym{mc}{MC}{Multi-Connectivity}
\newacronym{mcs}{MCS}{Modulation and Coding Scheme}
\newacronym{mec}{MEC}{Mobile Edge Cloud}
\newacronym{mi}{MI}{Mutual Information}
\newacronym{mimo}{MIMO}{Multiple Input, Multiple Output}
\newacronym{mmwave}{mmWave}{millimeter wave}
\newacronym{mptcp}{MPTCP}{Multipath TCP}
\newacronym{mr}{MR}{Maximum Rate}
\newacronym{mss}{MSS}{Maximum Segment Size}
\newacronym{mtd}{MTD}{Machine-Type Device}
\newacronym{mtu}{MTU}{Maximum Transmission Unit}
\newacronym{nfv}{NFV}{Network Function Virtualization}
\newacronym{nlos}{NLOS}{Non-Line-of-Sight}
\newacronym{nr}{NR}{New Radio}
\newacronym{ofdm}{OFDM}{Orthogonal Frequency Division Multiplexing}
\newacronym{pdcch}{PDCCH}{Physical Downlonk Control Channel}
\newacronym{pdcp}{PDCP}{Packet Data Convergence Protocol}
\newacronym{pdsch}{PDSCH}{Physical Downlink Shared Channel}
\newacronym{pdu}{PDU}{Packet Data Unit}
\newacronym{pf}{PF}{Proportional Fair}
\newacronym{pgw}{PGW}{Packet Gateway}
\newacronym{phy}{PHY}{Physical}
\newacronym{pbch}{PBCH}{Physical Broadcast Channel}
\newacronym[plural=\gls{mme}s,firstplural=Mobility Management Entities (MMEs)]{mme}{MME}{Mobility Management Entity}
\newacronym{prb}{PRB}{Physical Resource Block}
\newacronym{pss}{PSS}{Primary Synchronization Signal}
\newacronym{pucch}{PUCCH}{Physical Uplink Control Channel}
\newacronym{pusch}{PUSCH}{Physical Uplink Shared Channel}
\newacronym{rach}{RACH}{Random Access Channel}
\newacronym{ran}{RAN}{Radio Access Network}
\newacronym{red}{RED}{Random Early Detection}
\newacronym{rf}{RF}{Radio Frequency}
\newacronym{rlc}{RLC}{Radio Link Control}
\newacronym{rlf}{RLF}{Radio Link Failure}
\newacronym{rrc}{RRC}{Radio Resource Control}
\newacronym{rrm}{RRM}{Radio Resource Management}
\newacronym{rr}{RR}{Round Robin}
\newacronym{rs}{RS}{Remote Server}
\newacronym{rsrp}{RSRP}{Reference Signal Received Power}
\newacronym{rss}{RSS}{Received Signal Strength}
\newacronym{rtt}{RTT}{Round Trip Time}
\newacronym{rw}{RW}{Receive Window}
\newacronym{rx}{RX}{Receiver}
\newacronym{sa}{SA}{standalone}
\newacronym{sack}{SACK}{Selective Acknowledgment}
\newacronym{sap}{SAP}{Service Access Point}
\newacronym{sch}{SCH}{Secondary Cell Handover}
\newacronym{scoot}{SCOOT}{Split Cycle Offset Optimization Technique}
\newacronym{sdma}{SDMA}{Spatial Division Multiple Access}
\newacronym{sinr}{SINR}{Signal to Interference plus Noise Ratio}
\newacronym{sm}{SM}{Saturation Mode}
\newacronym{snr}{SNR}{Signal-to-Noise-Ratio}
\newacronym{son}{SON}{Self-Organizing Network}
\newacronym{ss}{SS}{Synchronization Signal}
\newacronym{srs}{SRS}{Sounding Reference Signal}
\newacronym{sss}{SSS}{Secondary Synchronization Signal}
\newacronym{tb}{TB}{Transport Block}
\newacronym{tcp}{TCP}{Transmission Control Protocol}
\newacronym{tdd}{TDD}{Time Division Duplexing}
\newacronym{tdma}{TDMA}{Time Division Multiple Access}
\newacronym{tfl}{TfL}{Transport for London}
\newacronym{tm}{TM}{Transparent Mode}
\newacronym{trp}{TRP}{Transmitter Receiver Pair}
\newacronym{tti}{TTI}{Transmission Time Interval}
\newacronym{ttt}{TTT}{Time-to-Trigger}
\newacronym{tx}{TX}{Transmitter}
\newacronym{ue}{UE}{User Equipment}
\newacronym{ul}{UL}{Uplink}
\newacronym{uml}{UML}{Unified Modeling Language}
\newacronym{um}{UM}{Unacknowledged Mode}
\newacronym{utc}{UTC}{Urban Traffic Control}
\newacronym{vm}{VM}{Virtual Machine}
\newacronym{rsrq}{RSRQ}{Reference Signal Received Quality}
\newacronym{rssi}{RSSI}{Received Signal Strength Indicator}
\newacronym{crs}{CRS}{Cell Reference Signal}
\newacronym{nsa}{NSA}{Non Stand Alone}
\newacronym{mrdc}{MR-DC}{Multi \gls{rat} \gls{dc}}
\newacronym{endc}{EN-DC}{E-UTRAN-\gls{nr} \gls{dc}}
\newacronym{5gc}{5GC}{5G Core}
\newacronym{si}{SI}{Study Item}
\newacronym{iab}{IAB}{Integrated Access and Backhaul}
\newacronym{wf}{WF}{Wired-first}
\newacronym{hqf}{HQF}{Highest-quality-first}
\newacronym{pa}{PA}{Position-aware}
\newacronym{mlr}{MLR}{Maximum-local-rate}
\newacronym{wbf}{WBF}{Wired Bias Function}
\newacronym{mib}{MIB}{Master Information Block}
\newacronym{sib}{SIB}{Secondary Information Block}
\newacronym{kpi}{KPI}{Key Performance Indicator}
\newacronym{ppp}{PPP}{Poisson Point Process}
\newacronym{gtp}{GTP}{GPRS Tunneling Protocol}
\newacronym{amf}{AMF}{Access and Mobility Management Function}
\tikzstyle{startstop} = [rectangle, rounded corners, minimum width=2cm, minimum height=0.5cm,text centered, draw=black]
\tikzstyle{io} = [trapezium, trapezium left angle=70, trapezium right angle=110, minimum width=3cm, minimum height=1cm, text centered, draw=black]
\tikzstyle{process} = [rectangle, minimum width=2cm, minimum height=0.5cm, text centered, draw=black, alignb=center]
\tikzstyle{decision} = [ellipse, minimum width=2cm, minimum height=1cm, text centered, draw=black]
\tikzstyle{arrow} = [thick,<->,>=stealth]
\tikzstyle{line} = [thick,>=stealth]
\tikzstyle{darrow} = [thick,<->,>=stealth,dashed]
\tikzstyle{sarrow} = [thick,->,>=stealth]
\tikzstyle{larrow} = [line width=0.1mm,dashdotted,->,>=stealth]

\makeatletter
\def\grd@save@target#1{%
  \def\grd@target{#1}}
\def\grd@save@start#1{%
  \def\grd@start{#1}}
\tikzset{
  grid with coordinates/.style={
    to path={%
      \pgfextra{%
        \edef\grd@@target{(\tikztotarget)}%
        \tikz@scan@one@point\grd@save@target\grd@@target\relax
        \edef\grd@@start{(\tikztostart)}%
        \tikz@scan@one@point\grd@save@start\grd@@start\relax
        \draw[minor help lines] (\tikztostart) grid (\tikztotarget);
        \draw[major help lines] (\tikztostart) grid (\tikztotarget);
        \grd@start
        \pgfmathsetmacro{\grd@xa}{\the\pgf@x/1cm}
        \pgfmathsetmacro{\grd@ya}{\the\pgf@y/1cm}
        \grd@target
        \pgfmathsetmacro{\grd@xb}{\the\pgf@x/1cm}
        \pgfmathsetmacro{\grd@yb}{\the\pgf@y/1cm}
        \pgfmathsetmacro{\grd@xc}{\grd@xa + \pgfkeysvalueof{/tikz/grid with coordinates/major step x}}
        \pgfmathsetmacro{\grd@yc}{\grd@ya + \pgfkeysvalueof{/tikz/grid with coordinates/major step y}}
        \foreach \x in {\grd@xa,\grd@xc,...,\grd@xb}
        \node[anchor=north] at (\x,\grd@ya) {\pgfmathprintnumber{\x}};
        \foreach \y in {\grd@ya,\grd@yc,...,\grd@yb}
        \node[anchor=east] at (\grd@xa,\y) {\pgfmathprintnumber{\y}};
      }
    }
  },
  minor help lines/.style={
    help lines,
    gray,
    line cap =round,
    xstep=\pgfkeysvalueof{/tikz/grid with coordinates/minor step x},
    ystep=\pgfkeysvalueof{/tikz/grid with coordinates/minor step y}
  },
  major help lines/.style={
    help lines,
    line cap =round,
    line width=\pgfkeysvalueof{/tikz/grid with coordinates/major line width},
    xstep=\pgfkeysvalueof{/tikz/grid with coordinates/major step x},
    ystep=\pgfkeysvalueof{/tikz/grid with coordinates/major step y}
  },
  grid with coordinates/.cd,
  minor step x/.initial=.5,
  minor step y/.initial=.2,
  major step x/.initial=1,
  major step y/.initial=1,
  major line width/.initial=1pt,
}
\makeatother

\definecolor{desireRed}{RGB}{230,57,60}%
\definecolor{darkPurple}{RGB}{59,31,43}%
\definecolor{springGreen}{RGB}{37,223,145}%
\definecolor{queenBlue}{RGB}{69,123,157}%
\definecolor{spaceCadet}{RGB}{29,53,87}%

\makeglossaries

\begin{document}
	
\title{End-to-End Simulation of\\Integrated Access and Backhaul at mmWaves}

\author{\IEEEauthorblockN{Michele Polese$^{\circ }$, Marco Giordani$^{\circ }$, Arnab Roy$^{\dagger }$, Sanjay Goyal$^{\dagger }$, Douglas Castor$^{\dagger }$, Michele Zorzi$^{\circ }$}
\IEEEauthorblockA{
\small email:\texttt{\{polesemi,giordani,zorzi\}@dei.unipd.it, }
\texttt{\small \{name.surname\}@interdigital.com}\\
$^{\circ }$\small Consorzio Futuro in Ricerca (CFR) and  University of Padova, Italy\\
$^{\dagger }$\small InterDigital Communications, Inc., USA\\}}

\flushbottom

\maketitle

\glsunset{nr}

\begin{abstract}
Recently, the millimeter wave (mmWave) bands have been investigated as a means to support the foreseen extreme data rate demands of next-generation cellular networks (5G).
However, in order to overcome the  severe isotropic path loss and the harsh propagation experienced at such high frequencies, a dense base station deployment is required, which may be infeasible because of the unavailability of fiber drops to provide wired backhauling. To address this challenge, the 3GPP is investigating the concept of Integrated Access and Backhaul (IAB), i.e., the possibility of providing  wireless backhaul to the mobile terminals. 
In this paper, we (i) extend the capabilities of the existing mmWave module for ns-3 to support advanced IAB functionalities, and (ii) evaluate the end-to-end performance of the IAB architecture through system-level full-stack simulations in terms of experienced throughput and communication latency.
We finally provide guidelines on how to design optimal wireless backhaul solutions in the presence of resource-constrained and traffic-congested mmWave scenarios.
\end{abstract}


\begin{picture}(0,0)(0,-380)
\put(0,0){
\put(0,0){\footnotesize This paper was accepted for publication at the 2018 IEEE International Workshop on Computer-Aided Modeling} 
\put(0,-10){\footnotesize Analysis and Design of Communication Links and Networks (CAMAD), September 2018, Barcelona, Spain.}}
\end{picture}

\section{Introduction}

The \gls{3gpp}
has recently started investigating new \glspl{rat} as enablers for the performance requirements of next-generation wireless systems, e.g., the so-called \gls{5g} network architecture~\cite{38300}.
\gls{5g} will be characterized by very stringent requirements in terms of latency, jitter and reliability, and is expected to provide the users with unprecedented data rates (e.g., on the order of gigabits per second~\cite{shafi2017deployment}), in line with the mobile data traffic predictions for 2020 and beyond~\cite{cisco2017}.
In this context, the \gls{mmwave} bands above 10 GHz  will play a key role, thanks to the very large bandwidths available at such high frequencies (up to 400 MHz per carrier, according to the latest \gls{3gpp} \gls{nr} specifications~\cite{38802}).
Moreover, the small size of antennas at
\glspl{mmwave} makes it practical to build very large antenna
arrays and obtain high beamforming gains, thereby overcoming the high propagation loss at such high frequencies and increasing the spectral efficiency.
On the other hand, \gls{mmwave} signals do not penetrate most
solid materials and are subject to high signal attenuation and
reflection, thus limiting the communication range of the \gls{mmwave} infrastructures. 
Although such an effect can be partially alleviated by configuring directional transmissions, a tracking mechanism is required to maintain the alignment of the transmitter and receiver beams, an operation that may increase the system overhead and lead to throughput degradation~\cite{rappaport2013millimeter}.

The combination of the high propagation loss and the blockage phenomenon calls for a high-density deployment of \glspl{gnb} (i.e., the base stations in NR terminology), to guarantee \gls{los} links at any given time and decrease the outage probability~\cite{rangan2017potentials}.
In such a deployment, providing wired backhaul to each of the \glspl{gnb} will be inevitably costly for network operators and, as a result, more economically sustainable solutions have been recently investigated by the 3GPP as part of the \gls{iab} \gls{si}~\cite{iabsi2017} for \gls{nr}. 
The idea  is to have wireless backhaul~\cite{dhillon2015wireless}, i.e., a fraction of  \glspl{gnb} with  traditional fiber-like backhaul capabilities and the rest of the \glspl{gnb} connected to the fiber infrastructures wirelessly, possibly through multiple hops.

The performance of this deployment paradigm is typically assessed  in terms of hop count and bottleneck \gls{snr}, e.g., in~\cite{polese2018distributed,kulkarni2018many}, which, however, prevents the comprehensive evaluation of the end-to-end performance of the different backhaul approaches.
In this regard, discrete-event network simulators enable full-stack simulation of complex and realistic scenarios, and therefore represent a viable tool for the setup of more accurate system-level~analysis.

In this paper, we therefore present the extension with the \gls{iab} features of the mmWave module for ns-3, which already models the mmWave channel and the \gls{phy} and \gls{mac} layers of the mmWave protocol stack~\cite{mezzavilla2018endtoend}. This extension can support both single- and multi-hop deployments and autonomous network configuration, and features a detailed 3GPP-like protocol stack implementation. Moreover, new scheduling mechanisms have been developed in order to support the sharing of access and backhaul resources.

We also evaluate the performance of the proposed \gls{iab} architecture in an end-to-end environment in terms of experienced throughput and latency, considering realistic traffic models. Our preliminary results demonstrate that the configuration of a wireless backhaul deployment has the potential to increase the overall network throughput as well as to reduce the communication latency in case of congested networks, and therefore represents a practical solution in future \gls{mmwave} networks.
Our simulator can be used for the 
design of enhanced multi-hop routing strategies and scheduling algorithms, and for the determination of the best strategy for the deployment of a wireless backhaul~solution.

The rest of the paper is organized as follows. Sec.~\ref{sec:iabsi} describes the characteristics of the 3GPP \gls{si} on \gls{iab} and the potential of this solution for \gls{nr} deployments. Sec.~\ref{sec:model} presents the implementation of the \gls{iab} features in ns-3,  while Sec.~\ref{sec:perf_eval} provides a preliminary evaluation of the end-to-end  performance of the \gls{iab} nodes in a realistic \gls{mmwave} scenario. Finally, Sec.~\ref{sec:concl} concludes the paper and discusses possible extensions of this work.

\section{State of The Art on \\ Integrated Access and Backhaul}
\label{sec:iabsi}

Research on  wireless backhaul solutions  has been carried out in the past at frequencies below 6 GHz, e.g., in the WLAN domain~\cite{gambiroza2004multihop} and as part of the \gls{lte} standardization activity with a single wireless backhaul hop~\cite{36826}.
The practical implementation of wireless multi-hop  networks, however, never really turned into a commercial deployment due to practical limitations including, but not limited to, (i) scalability issues~\cite{dhillon2015wireless}, (ii) the scheduling constraints between hops~\cite{sikora2006bandwidth}, and (iii) the large overhead for maintaining multi-hop routes~\cite{andrews2008rethinking}.

Nonetheless, with the recent advancements in \gls{mmwave} communication and leveraging 
highly directional beamforming, the integration of wireless backhaul and radio access is being considered as a promising solution for \gls{5g} cellular networks. 
In~\cite{singh2015tractable}, the authors demonstrated that the noise-limited nature of large-bandwidth mmWave networks offer
interference isolation, thereby providing an opportunity to incorporate
self-backhauling in a mesh
small-cell deployment without significant throughput degradation.
Paper~\cite{ge2014wireless}  showed that wireless backhaul over mmWave links can meet the expected increase in mobile traffic demands, while paper~\cite{mesodiakaki2016energy} evaluates the energy efficiency of mmWave backhaul at different frequencies. 
The authors in~\cite{saha2018integrated} further evaluated the performance of the integration between access and backhaul and determined the  maximum total network load that can be supported using the
\gls{iab}~architecture.

Along these lines, the 3GPP is also focusing on IAB for
3GPP NR \cite{iabsi2017}, to design an
advanced wireless relay which overcomes the limitations of
the traditional \gls{lte} implementation and makes it possible to flexibly deploy self-backhauled
NR base stations.
According to~\cite{iabsi2017}, \gls{nr} cellular networks with \gls{iab} functionalities will be characterized by (i) the possibility of using the \gls{mmwave} spectrum; (ii) the integration of the access and backhaul technologies, i.e.,  using the same spectral resources and
infrastructures to serve both mobile terminals
in access as well as the NR \glspl{gnb} in backhaul~\cite{38874}; (iii) a higher flexibility in terms of network deployment and configuration with respect to \gls{lte}, i.e., the possibility of deploying plug-and-play \gls{iab} nodes capable of self-configuring and self-optimizing themselves~\cite{22261}.
According to~\cite{iabsi2017,22261}, 5G \gls{iab} relays will be used in both outdoor and indoor scenarios, possibly with multiple wireless hops, with the final goal of extending the coverage of cell-edge users, avoiding service unavailability, and increasing the efficiency of the resource allocation. Both in-band and out-of-band backhaul will be considered, with the first being a natural candidate for a tighter integration between access and backhaul. 

Despite its clear strengths, the design of IAB solutions in mmWave systems is a research challenge that is still largely unexplored.
Most of the existing literature 
does not consider a channel characterized by the full channel matrix,  nor realistic beamforming patterns.
Moreover, the prior art lacks considerations on the end-to-end performance of the self-backhauling architectures, which are in turn part of our original~contributions.

\section{IAB in ns-3 mmWave}
\label{sec:model}

The ns-3 mmWave module, described in~\cite{mezzavilla2018endtoend}, enables the simulation of end-to-end cellular networks at mmWave frequencies. 
It features a complete stack for \glspl{ue} and \glspl{gnb}, with a custom \gls{phy} layer, described in~\cite{dutta2017frame}, the 3GPP mmWave channel model and, thanks to the integration with ns-3, a complete implementation of the TCP/IP protocol stack.

As mentioned in the previous sections, \gls{iab} will be important for \gls{nr} ultra-dense mmWave deployments.\footnote{The 3GPP \gls{si} on IAB is still ongoing and is scheduled for completion as part of Release 16. We therefore do not preclude in the future to further extend the features of the ns-3 IAB module to make it fully compliant with the latests 3GPP specifications on this topic.} Therefore, in order to increase the realism and the modeling capabilities of the ns-3 mmWave module, we implemented an \gls{iab} framework that will be described in the following sections. It features a new ns-3 \texttt{NetDevice}, the \texttt{MmWaveIabNetDevice} with a dual stack for access and backhaul, an extension of the ns-3 mmWave module schedulers, and network procedures to support \gls{iab} nodes in a simulation scenario. Moreover, we simulate the wireless relaying of both data and control plane messages, in order to accurately model the \gls{iab} operations.

\begin{figure}
	\centering	
	\includegraphics[width=0.9\columnwidth]{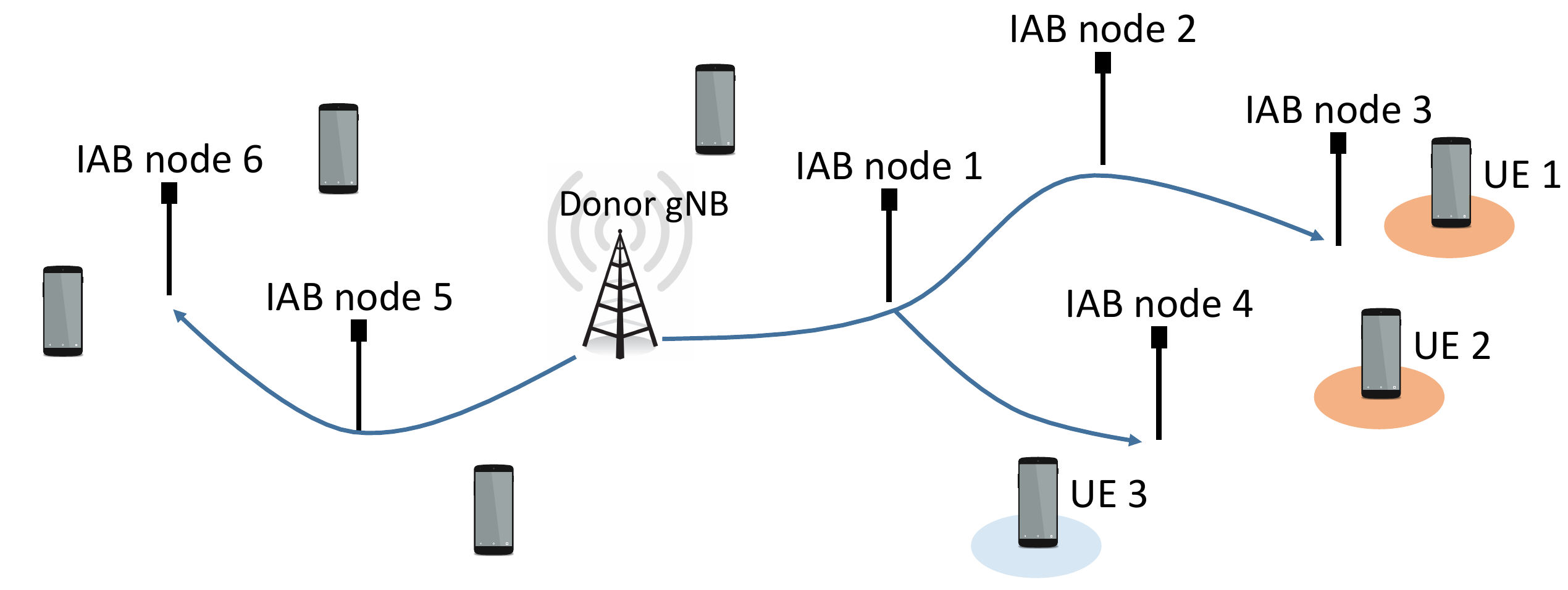}
	\caption{Example of \gls{iab} architecture, with a single donor and multiple downstream \gls{iab} nodes.}
	\label{fig:iab_arch}
\end{figure}

An example of \gls{iab} network that can be now supported by ns-3 is shown in Fig.~\ref{fig:iab_arch}. In particular, we consider a tree architecture, with the root being a donor \gls{gnb}, i.e., a base station with a wired connection to the core network. Therefore, this is not a traditional mesh architecture, which is used, for example, for random-access-based backhaul technologies such as IEEE 802.11~\cite{alicherry2005joint}, in which there is no strict parent/child relationship between network nodes. In a cellular context, it is necessary to define a tree structure because every communication is scheduled~\cite{38300}, i.e., the base station assigns specific time and frequency resources for downlink or uplink communication with any connected \gls{ue}. Therefore, given that the access and the backhaul share the same resources, then also communication between the \gls{gnb} and any \gls{iab} node must be scheduled. Notice that the connection between a parent and a child node can change with handover procedures\footnote{We will introduce support for this functionality in the next iteration of the module.}, for example if the link quality between them degrades because~of~blockage.

In the following paragraphs we will describe the protocol stack that is deployed in each \gls{iab} node, the scheduling mechanism, and how to set up a simulation with \gls{iab} features.

\subsection{IAB node}
\label{sec:iab}
\begin{figure}[t]
	\centering
	\includegraphics[width=.98\columnwidth]{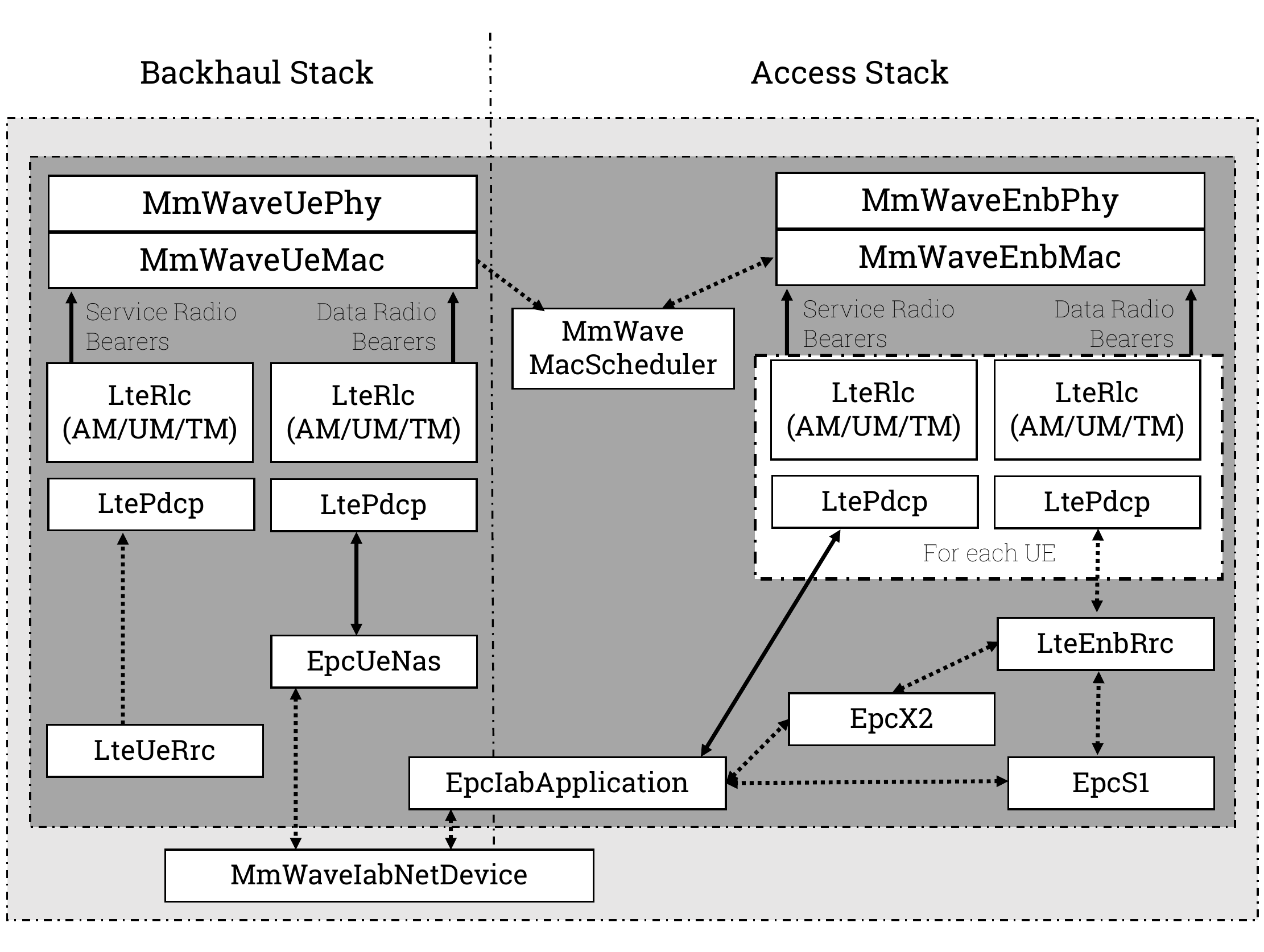}
	\caption{Protocol stack and organization of the ns-3 classes for an \gls{iab} node.}
	\label{fig:classes}
\end{figure}

As mentioned in~\cite{iabsi2017}, the \gls{iab} nodes should re-use the specifications for the access stack of \gls{nr} as much as possible. At the moment, there are a few protocol stacks being discussed in the 3GPP~\cite{iab2018architecture}. All of them, however, include \gls{phy}, \gls{mac} and \gls{rlc} layers, and differ because of the support of layer-2 (i.e., \gls{rlc} or \gls{pdcp}) or layer-3 relaying. Given the need for a flexible solution, able to adapt to the direction that the 3GPP will take, we implemented a light layer-3 relaying solution, i.e., each backhaul radio bearer is set up locally, and a middle layer above the \gls{pdcp} handles the forwarding of the packets from the access to the backhaul \glspl{pdcp}. Fig.~\ref{fig:classes} shows the protocol stack for an \gls{iab} node and the classes that model it.

The main novelties are the \texttt{MmWaveIabNetDevice} and the \texttt{EpcIabApplication} classes. The first is an extension of the ns-3 \texttt{NetDevice} class, and, similarly to the \texttt{NetDevice} implementations of the \gls{ue} and \gls{gnb}, holds pointers to all the objects that model the other layers of the protocol stack. Moreover, it is internally used in the ns-3 model to forward packets between an instance of the \texttt{EpcUeNas} class~in the backhaul stack and the \texttt{EpcIabApplication} in the access~stack.

The \texttt{EpcIabApplication}, instead, implements the main logic related to the control and data plane management in the \gls{iab} node. In particular, for the data plane, the \texttt{EpcIabApplication} class is in charge of applying the forwarding rules for local \glspl{ue}, i.e., those directly connected to the \gls{iab} node this class belongs to, and for remote \glspl{ue}, i.e., those connected to downstream \gls{iab} nodes. In this case, the traffic will be forwarded to the local bearer mapped to the downstream \gls{iab} device. More details on how the routing is performed will be given in Sec.~\ref{sec:proc}. 
This class is also responsible for the processing and forwarding of control packets for the interfaces toward the core network and the other neighboring \glspl{gnb}. When a control message is received on either the access or the backhaul interface, the \texttt{EpcIabApplication} checks if it is a local message, i.e., if the destination is the \gls{rrc} layer of the current \gls{iab} node, and, if this is the case, forwards the packet to the \gls{rrc}. Otherwise, as done in the data plane, the packets are relayed via one of the downstream \gls{iab} nodes. 

The other classes are the same as those used in the \gls{ue} protocol stack (for the backhaul) and \gls{gnb} protocol stack (for the access). The consequence is that, in the access, the \glspl{ue} in the scenario consider the \gls{iab} node as a normal \gls{gnb}, and, similarly, in the backhaul, the parent \glspl{gnb} and/or \gls{iab} nodes consider the \gls{iab} child as a \gls{ue}. Therefore, there is no need to adapt the \gls{ue} and \gls{gnb} ns-3 implementations to support the \gls{iab} feature. The only change is the extension of the \gls{gnb} schedulers, to support the multiplexing of access and backhaul in the same resources, and the introduction of a new interface between the access and backhaul \gls{mac} layers. These extensions will be described in Sec.~\ref{sec:mac}. Nonetheless, additional enhancements can be introduced in future releases, to improve the overall performance of the \gls{iab} protocol stack and track the 3GPP \gls{si} and specifications on \gls{iab}.


\subsection{Single- and multi-hop control procedures}
\label{sec:proc}
Given that the 3GPP is still considering \gls{iab} as an \gls{si}, there are no standard specifications yet on control procedures to support \gls{iab} networks. Nonetheless, the \gls{si}~\cite{iabsi2017} specifies that both single- and multi-hop topologies should be considered, and that the \gls{iab} node should be able to autonomously connect to the network, adapt the access and backhaul resource partitioning and, eventually, independently update the parent node in case of blockage. 
All these features require specific control procedures, and, given the high level of detail of the ns-3 model, we implemented a number of realistic control procedures, which involve an exchange of messages on the wireless backhaul links to set up and automatically configure the network. 
These can be easily updated to implement different network procedures that the 3GPP may specify in the future.

In particular, we assume that the parent \gls{iab} node for a backhaul link terminates the NG control interface to the core network (i.e., the NR equivalent of the \gls{lte} S1 interface)~\cite{38401}, and that it takes care of forwarding the control messages towards the network servers that host the \gls{amf}. Moreover, the \gls{iab} node has a similar role with respect to the \glspl{ue} connected to it, as would happen with a traditional wired \gls{gnb}. 
Thanks to this design, the differences with respect to the 3GPP specifications for the access stack are minimized. 
This configuration makes it possible to seamlessly support both single- and multi-hop deployments, given that the architecture of the upstream portion of the network is transparent to each \gls{iab} node, which will simply relay all of its packets to the parent. Furthermore, for the purpose of packet transport in the backhaul network, we exploit \gls{gtp} tunnels from each \gls{iab} node to the relevant element in the core network (i.e., the server with control functions or the packet gateway). Each data bearer of all the \glspl{ue} (and \gls{iab} nodes, for the backhaul part) is associated with a unique tunneling ID, and all the packets sent on backhaul links will be associated with a \gls{gtp} header carrying that ID.

We also implemented realistic autonomous access and configuration procedures for the IAB nodes. When the \gls{iab} selects its parent node during the \gls{ia} procedure\footnote{Initial access is the procedure by which a mobile terminal establishes an initial physical link connection with a cell, a necessary step to access the network. For a complete overview of the most relevant works on IA for 3GPP NR scenario we refer the reader to \cite{giordani2018initial,giordani2018tutorial}.}, the parent sends an initial message to the \gls{amf}, which will reply with the configuration for the backhaul bearer between the \gls{iab} node and its parent. These messages will be relayed by all the \gls{iab} nodes in the path between the parent and the donor \gls{gnb}, and each of them will register the presence of an additional downstream \gls{iab} device. Notice that there may be multiple \gls{iab} children for each parent, therefore the parent has to match the new downstream node to the correct child to correctly route the other control and data packets. 

For the \glspl{ue}, there is no difference between a wireless relay and a \gls{gnb} with a wired connection to the core network. Therefore, the \gls{ue}'s \gls{ia} procedure does not change, and the \gls{iab} node will take care of forwarding the relevant control messages to the \gls{amf} and the other network functions involved in the \gls{ia}. Moreover, the upstream relays and the donor \gls{gnb} will exploit the control messages for the \gls{ue}'s \gls{ia} to associate to each \gls{iab} bearer the total number of downstream \glspl{ue}. For example, by considering the deployment in Fig.~\ref{fig:iab_arch}, if \glspl{ue} 1 and 2 connect to \gls{iab} node 3, and \gls{ue} 3 connects to \gls{iab} node 4, then \gls{iab} node 1 will know that the backhaul bearer towards \gls{iab} node 2 will carry the traffic for 2 \glspl{ue}, and that towards node 4 will account for a single \gls{ue}. This information could be exploited by advanced \gls{iab} \gls{mac} schedulers.
Finally, during the \gls{ue} \gls{ia} procedure, each \gls{gnb} associates the \gls{gtp} tunneling ID of the bearers of downstream \glspl{ue} to a local \gls{iab} child, so that, when a backhaul packet is received, the \gls{gnb} uses the information  in the \gls{gtp} header to correctly route the~packet.

\subsection{Backhaul-aware dynamic scheduler}
\label{sec:mac}
The \gls{mac} and the associated scheduler are a key component in the design of scheduled wireless relay architectures in which the resources between the access and the backhaul are shared. In order to avoid self-interference between access and backhaul, indeed, there is a need to multiplex the two interfaces. In our implementation, we consider \gls{tdma}, but we plan to extend the support to spatial division multiplexing in future releases, to harness the directionality of mmWave communications. Moreover, the scheduler is usually not part of the 3GPP specifications, and, therefore, equipment vendors have the possibility of designing custom solutions in this domain. 

We opted for a distributed scheduling solution, in order to minimize the difference in the scheduling mechanism with respect to a traditional access-only scenario, and to limit the amount of control overhead that a centralized solution would require. Therefore, in the ns-3 mmWave \gls{iab} module, each \gls{gnb} (either wired or wireless) schedules the resources for its access interface (i.e., for both \glspl{ue} and \gls{iab} children) independently of the other \glspl{gnb}, as would happen in a traditional network without \gls{iab}. In a \gls{tdma} setup, however, the \gls{iab} node cannot schedule resources in the time and frequency slots already allocated to the backhaul by their parent. Therefore, if at time $t$ the relay has to perform a scheduling decision for subframe $t+\eta$, then it has to be already aware of the scheduling decision of its parent for $t+\eta$. Given a delay $\epsilon$ for the communication of scheduling information between the parent and the current relay, then the parent should perform its scheduling decisions for $t+\eta$ at time $t-\epsilon$. 

In order to efficiently address this issue, we implemented a \textit{look-ahead backhaul-aware scheduling} mechanism. The backhaul-aware component is given by a new interface between the access and the backhaul \gls{mac} layers. The backhaul \gls{mac} layer is seen as a \gls{ue} by the parent node, and thus will receive \glspl{dci} with the scheduling and modulation and coding scheme information for $\eta$ subframes in advance. Then, the backhaul \gls{mac} shares \gls{dci} with the scheduler of the \gls{iab} node (in the access stack), which registers the resources occupied by backhaul transmissions for the relevant subframe.

The look-ahead mechanism, additionally, makes it possible to adjust the value of $\eta$ according to the maximum number of downstream relaying hops $N$ from the current \gls{gnb} to the farthest \gls{iab} node: the \gls{gnb} schedules ahead by $\eta = N + 1$ subframes\footnote{The additional subframe with respect to $N$ is needed because the farthest \gls{iab} node (without \gls{iab} children) has to schedule its resources at least one subframe in advance, in order to transmit the \gls{dci} beforehand to its~\glspl{ue}}, and propagates this information with a \gls{dci} to the \glspl{ue} and \gls{iab} nodes connected to it. In turn, these \gls{iab} nodes will schedule ahead by at most $\eta = N$ subframes.
Each of them will consider the time and frequency resources allocated for their downlink or uplink backhaul transmission as busy, and will schedule access resources for their \glspl{ue} and, eventually, for \gls{iab} nodes in unallocated resources. For example, by considering Fig.~\ref{fig:iab_arch}, the maximum number of hops from the donor \gls{gnb} is 3. Therefore, the donor will schedule ahead by 4 subframes. On the other hand, \gls{iab} node 2 has a single hop to the farthest  relay, thus it will schedule ahead by 2 subframes. Notice that, thanks to the procedures introduced in Sec.~\ref{sec:proc}, there is no need to manually tune the $\eta$ parameter, which is automatically configured according to the structure of the \gls{iab} tree, and can be updated in case of variations in the architecture of the network.

\begin{figure}
	\centering
	\includegraphics[width=0.95\columnwidth]{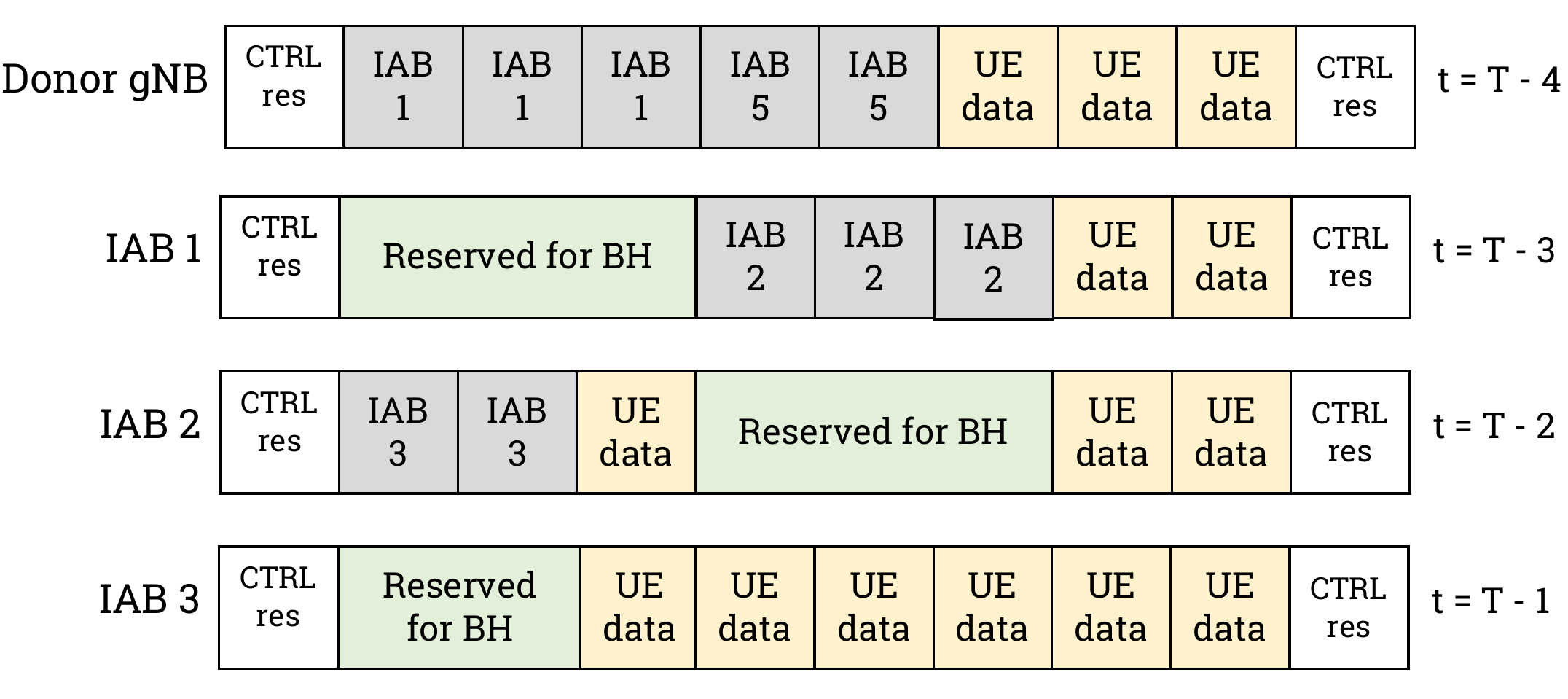}
	\caption{Example of resource allocation for time $T$ with a look-ahead scheduler at the donor \gls{gnb} and \gls{iab} nodes 1, 2 and 3 in the deployment of Fig.~\ref{fig:iab_arch}.}
	\label{fig:sched}
\end{figure}

We added the look-ahead and backhaul-aware capabilities to two of the ns-3 mmWave module schedulers, i.e., the \texttt{MmWaveFlexTtiMacScheduler} class, which models a \gls{rr} scheduler, and the \texttt{MmWaveFlexTtiPfMacScheduler} class, which implements a \gls{pf} scheduling algorithm. Moreover, in a \gls{tdma} setup, with shared resources between the access and the backhaul, it is important to make sure that the parent \gls{gnb} does not schedule all of the available resources to a single \gls{iab} node (e.g., if it is the only active terminal connected to the parent). Otherwise, the child \gls{iab} node would not be able to allocate any resource to the access. Therefore, we limit the maximum number of time and frequency resources that can be allocated to an \gls{iab} device to half of the total available~resources. 

An example of resource allocation is shown in Fig.~\ref{fig:sched}, where a total number of 10 time and frequency resources are dynamically allocated to access and backhaul links. 
In particular, we refer to the deployment in Fig.~\ref{fig:iab_arch}, and present a possible resource partitioning for the donor \gls{gnb}, \gls{iab} nodes 1, 2 and 3 and the \glspl{ue} connected to these \glspl{gnb}. 
As can be seen, each \gls{iab} node does not allocate access transmission in the resources reserved for its backhaul, but can exploit all of the other resources for communication with other relays and the \glspl{ue}, including those allocated by one of the upstream nodes to other backhaul links. 
While in general this may increase the interference, it must be noticed that, at mmWave frequencies, the large antenna arrays that can be built and the resulting directional transmissions that can be established have the potential to provide increased spatial reuse and isolation, thereby guaranteeing reduced interference~\cite{rebato2018study}. 
Moreover, interference-aware schedulers can be designed and tested with the simulator. Finally, it is possible to update the allocation on-the-fly, to dynamically adapt to changed channel conditions and traffic requirements from the different connected terminals. 

Fig.~\ref{fig:sched}, however, also highlights one of the main bottlenecks of an \gls{iab} architecture, i.e., the fact that the donor \gls{gnb} needs to serve not only its own users, but also all the downstream relays, carrying traffic from many other \glspl{ue}. On one hand, the amount of data that can be exchanged on a backhaul link in each time and frequency resource is generally higher than the equivalent for a \gls{gnb}-\gls{ue} link, thus the backhaul will probably require fewer resources. Indeed, the backhaul link between two \glspl{gnb} has usually a better quality than that between a \gls{gnb} and a \gls{ue}, given that a backhaul link is expected to be in \gls{los}, and that a larger number of antennas can be deployed in a relay than in a \gls{ue}. On the other hand, the scalability of an \gls{iab} deployment has an intrinsic limitation due to the resource sharing between the access and the backhaul link. Therefore, efficient scheduling algorithms will be key for high-performance \gls{iab} networks. This makes the ns-3 mmWave module with the \gls{iab} integration a valuable platform for researchers interested in \gls{iab} networks, given that it offers a lean interface to the scheduler implementations, which can be easily extended to test new \gls{iab} scheduling paradigms in realistic end-to-end scenarios.

\subsection{Simulation setup}
\label{sec:sim_setup}
The setup of a simulation with the \gls{iab} feature resembles that of a simulation with traditional wired-only backhaul. An extensive description of how to configure an ns-3 simulation script for the mmWave module is provided in the tutorial in~\cite{mezzavilla2018endtoend}.
We added two auxiliary methods in the \texttt{MmWaveHelper} class, which hides from the ns-3 user much of the complexity related to the configuration of the mmWave \gls{ran} and core network. Similarly to the methods used to set up \glspl{ue} and \glspl{gnb}, the \texttt{InstallIabDevice} method returns a \texttt{NetDevice} properly configured, with the stack described in Fig.~\ref{fig:classes}. 

The initial attachment of each \gls{iab} node to its parent \gls{gnb} is performed by the methods \texttt{AttachIabToClosestWiredEnb} or \texttt{Attach\-IabToBestNodeHQF}. The latter scans the signal quality of the available \gls{iab} nodes or wired donors, and selects that with the highest \gls{snr}. Moreover, it avoids the creation of loops in the network tree. These helper methods, moreover, automatically register the new \gls{iab} nodes to the control entities in the core network, and define the default radio bearer that will be used for the backhaul link. Finally, by default, the \glspl{ue} in the ns-3 mmWave module perform the initial attachment as soon as the simulation starts, i.e., at simulation time $t_s=0$. Therefore, we added the \texttt{AttachToClosestEnbWithDelay} method that delays by $D$ seconds the initial attachment of \glspl{ue} to the chosen \glspl{gnb}, either wired or wireless. This method can be used to let the \glspl{ue} perform \gls{ia} only after the \gls{iab} nodes have completed their \gls{ia} and backhaul bearer setup.

\section{Example results}
\label{sec:perf_eval}

In this section, we validate the implementation of the IAB features for the ns-3 mmWave module through simulations.
We  illustrate some preliminary results related to the coverage performance of an \gls{iab} deployment in a mmWave environment. The considered scenario is a Manhattan grid, with blocks of 50~m for each side, and with 10~m between each block, for a total area of 0.053~km$^2$. A \gls{gnb} with a wired connection to the core network is placed at the center of the scenario, while the number of \gls{iab} nodes (i.e., \glspl{gnb} with wireless backhaul functionalities) varies from 0 to 4. 
The IAB nodes are one block in each direction away from the donor (i.e., at a distance of 85~m), and they are in \gls{los} (e.g., placed on the building rooftops). Each relay directly connects to the wired donor wirelessly, thus this scenario only considers single-hop transmissions.\footnote{Although our simulator enables multi-hop relaying operations, for the tractability of the simulation in this paper we only focus on  single-hop transmissions, and we leave the analysis of the multi-hop architecture  as part of our future work.}
40 users are randomly placed outdoors using the new ns-3 \texttt{OutdoorPositionAllocator} method, and connect to the closest  \gls{gnb}, either wired or wireless. Each \gls{ue} downloads content from a remote server
at a constant rate $R =\{28, 224\}$ Mbit/s using UDP as the transport protocol. These two different source rates are used to test the network in different congestion conditions.
Finally, the \gls{mac} layer performs \gls{harq} retransmissions, and the \gls{rlc} layer uses the \gls{am} to provide additional reliability. The scheduler is Round Robin, with the look-ahead backhaul-aware mechanisms described in Sec.~\ref{sec:mac}.
The other simulation parameters are~in~Table~\ref{table:params}.

\begin{table}[t]
  \centering
  \small
  \begin{tabular}{@{}ll@{}}
    \toprule
    Parameter & Value \\
    \midrule
    mmWave carrier frequency & 28 GHz \\
    mmWave bandwidth & 1 GHz \\
    3GPP Channel Scenario & Urban Micro \\
    mmWave max PHY rate & 3.2 Gbit/s \\
    \gls{mac} scheduler & Round Robin \\
    Subframe duration & 1 ms\\
    Donor \gls{gnb} to remote server latency & 11 ms \\
    RLC buffer size $B_{RLC}$ for \glspl{ue} & 10 MB \\
    RLC buffer size $B_{RLC}$ for \gls{iab} nodes & 40 MB \\
    RLC AM reordering timer & 2 ms \\
    UDP rate $R$ & $\{28, 224\}$ Mbit/s \\
    UDP packet size & 1400 byte \\
    Number of independent simulation runs & 50 \\
    \bottomrule
  \end{tabular}
  \caption{Simulation parameters}
  \label{table:params}
\end{table}

We consider two different end-to-end metrics, i.e., the experienced throughput and the application-layer latency averaged over multiple independent runs. Fig.~\ref{fig:th} investigates three different throughput values for different source rates $R$ and varying the number of IAB relays. We observe that, for the low source rate scenario (i.e., $R=28$ Mbit/s), the total throughput remains almost constant, while, in the congested scenario (i.e.,  $R=224$ Mbit/s) the rate progressively increases with the number of relays. 
This shows that, in the considered Manhattan scenario, the  relays  extend the area in which the mobile terminals can benefit from the coverage of their serving infrastructures and, in particular, have the potential to improve the quality of the access link between the cell-edge users and the donor \gls{gnb}, thereby guaranteeing higher capacity.

\begin{figure}[t]
  \setlength\fwidth{0.75\columnwidth}
  \setlength\fheight{0.4\columnwidth}
%
%
\definecolor{mycolor1}{rgb}{0.28101,0.32276,0.95789}%
\definecolor{mycolor2}{rgb}{0.34064,0.80080,0.47886}%
\begin{tikzpicture}
\pgfplotsset{every tick label/.append style={font=\scriptsize}}

\begin{axis}[%
width=1.03\fwidth,
height=\fheight,
at={(0\fwidth,0\fheight)},
scale only axis,
unbounded coords=jump,
xmin=0,
xmax=4,
xlabel style={font=\scriptsize\color{white!15!black}},
xlabel={Number of relays},
ymin=0,
ymax=350,
ylabel style={font=\scriptsize\color{white!15!black}},
ylabel={Average UDP latency [ms]},
axis background/.style={fill=white},
xmajorgrids,
ymajorgrids,
legend style={font=\scriptsize, at={(0.44, 1.03)}, anchor=south, legend cell align=left, align=left, draw=white!15!black},
legend columns=2,
]
\addplot [color=mycolor1, mark=+, mark options={solid, mycolor1}]
  table[row sep=crcr]{%
0	292.007188945677\\
1	272.207096427852\\
2	243.90010953837\\
3	162.926240594003\\
4	30.946797958039\\
};
\addlegendentry{Donor \gls{gnb} \glspl{ue}, $R=224$ Mbit/s}

\addplot [color=mycolor2, mark=x, mark options={solid, mycolor2}]
  table[row sep=crcr]{%
0	57.3280252927493\\
1	51.8961694751781\\
2	47.9762340242121\\
3	54.6764028331412\\
4	16.9759971303606\\
};
\addlegendentry{Donor \gls{gnb} \glspl{ue}, $R=28$ Mbit/s}

\addplot [color=mycolor1, dashed, mark=+, mark options={solid, mycolor1}]
  table[row sep=crcr]{%
0	nan\\
1	345.822777950651\\
2	341.845398757932\\
3	328.839359022452\\
4	310.21585345025\\
};
\addlegendentry{\gls{iab} nodes \glspl{ue}, $R=224$ Mbit/s}

\addplot [color=mycolor2, dashed, mark=x, mark options={solid, mycolor2}]
  table[row sep=crcr]{%
0	nan\\
1	176.24985151644\\
2	85.5132578181366\\
3	77.4032541189312\\
4	81.7506468823201\\
};
\addlegendentry{\gls{iab} nodes \glspl{ue}, $R=28$ Mbit/s}

\addplot [color=mycolor1, dashdotted, line width=1.5pt, mark=+, mark options={solid, mycolor1}]
  table[row sep=crcr]{%
0	292.007188945677\\
1	309.014937189251\\
2	309.196969018078\\
3	287.36107941534\\
4	254.362042351808\\
};
\addlegendentry{All \glspl{ue}, $R=224$ Mbit/s}


\addplot [color=mycolor2, dashdotted, line width=1.5pt, mark=x, mark options={solid, mycolor2}]
  table[row sep=crcr]{%
0	57.3280252927493\\
1	114.073010495809\\
2	73.0009165534951\\
3	71.7215412974837\\
4	68.7957169319282\\
};
\addlegendentry{All \glspl{ue}, $R=28$ Mbit/s}

\end{axis}
\end{tikzpicture}%
  \caption{Average end-to-end latency for different source rate $R$ and number of relays. We report the average latency considering all the \glspl{ue}, or only those connected to the wired \gls{gnb} or wireless relays. The dotted black line represents the average latency of the configuration with 0 relays.}
  \label{fig:lat}
\end{figure}
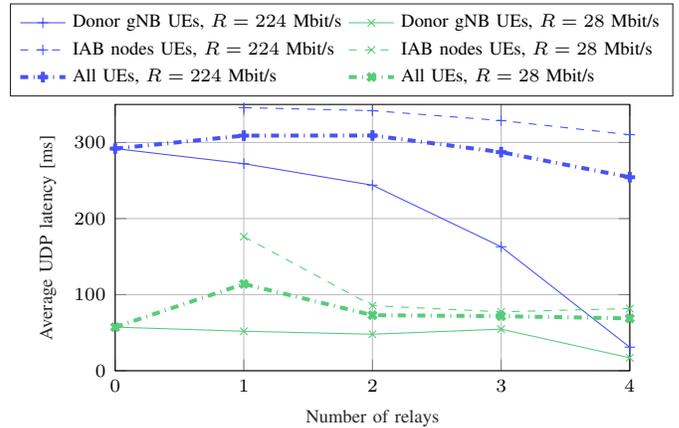

The average latency is shown in Fig.~\ref{fig:lat}. 
We see that, in a Manhattan grid scenario, the average latency of the \glspl{ue} directly connected to the wired \gls{gnb} decreases as a result of increasing the number of wireless relays.
Indeed, if the relays are used, the wired gNB will serve fewer users, i.e., those with the best channel quality, and will avoid allocating resources to cell-edge users which, generally, require a high number of \gls{harq} and \gls{rlc} retransmissions.
Although these  benefits are particularly evident in the $R = 224$ Mbit/s case, a latency improvement is also observed for the non-congested scenario (i.e., $R=28$ Mbit/s) when four relays are deployed.

\begin{figure}[t]
  \setlength\fwidth{0.75\columnwidth}
  \setlength\fheight{0.4\columnwidth}
%
%
\definecolor{mycolor1}{rgb}{0.28101,0.32276,0.95789}%
\definecolor{mycolor2}{rgb}{0.34064,0.80080,0.47886}%
\begin{tikzpicture}
\pgfplotsset{every tick label/.append style={font=\scriptsize}}

\begin{axis}[%
width=1.03\fwidth,
height=\fheight,
at={(0\fwidth,0\fheight)},
scale only axis,
xmin=0,
xmax=4,
xlabel style={font=\scriptsize\color{white!15!black}},
xlabel={Number of relays},
ymin=0,
ymax=2500,
ylabel style={font=\scriptsize\color{white!15!black}},
ylabel={Sum UDP throughput [Mbit/s]},
axis background/.style={fill=white},
xmajorgrids,
ymajorgrids,
yticklabel shift=-2pt,
ylabel shift=-4pt,
legend style={font=\scriptsize, at={(0.44, 1.03)}, anchor=south, legend cell align=left, align=left, draw=white!15!black},
legend columns=2,
]
\addplot [color=mycolor1, mark=+, mark options={solid, mycolor1}]
  table[row sep=crcr]{%
0	1746.47673469388\\
1	1611.22976\\
2	1634.86688\\
3	1578.88096\\
4	1354.7488\\
};
\addlegendentry{Donor \gls{gnb} \glspl{ue}, $R=224$ Mbit/s}

\addplot [color=mycolor2, mark=x, mark options={solid, mycolor2}]
  table[row sep=crcr]{%
0	964.730448979592\\
1	716.31968\\
2	551.21824\\
3	318.36064\\
4	177.01024\\
};
\addlegendentry{Donor \gls{gnb} \glspl{ue}, $R=28$ Mbit/s}

\addplot [color=mycolor1, dashed, mark=+, mark options={solid, mycolor1}]
  table[row sep=crcr]{%
1	84.5376\\
2	219.5664\\
3	532.61856\\
4	958.24608\\
};
\addlegendentry{\gls{iab} nodes \glspl{ue}, $R=224$ Mbit/s}

\addplot [color=mycolor2, dashed, mark=x, mark options={solid, mycolor2}]
  table[row sep=crcr]{%
1	153.9344\\
2	380.00768\\
3	594.15424\\
4	728.45504\\
};
\addlegendentry{\gls{iab} nodes \glspl{ue}, $R=28$ Mbit/s}

\addplot [color=mycolor1, dashdotted, mark=+, mark options={solid, mycolor1}]
  table[row sep=crcr]{%
0	1746.47673469388\\
1	1695.76736\\
2	1854.43328\\
3	2111.49952\\
4	2312.99488\\
};
\addlegendentry{All \glspl{ue}, $R=224$ Mbit/s}

\addplot [color=mycolor2, dashdotted, mark=x, mark options={solid, mycolor2}]
  table[row sep=crcr]{%
0	964.730448979592\\
1	870.25408\\
2	931.22592\\
3	912.51488\\
4	905.46528\\
};
\addlegendentry{All \glspl{ue}, $R=28$ Mbit/s}

\end{axis}
\end{tikzpicture}%
  \caption{Sum end-to-end throughput for different source rate $R$ and number of relays. The total throughput is the sum of the throughput of all the users, while the wired (or \gls{iab} nodes) sum throughput refers to the aggregate throughput of \glspl{ue} connected to the donor (or the relays, respectively).}
  \label{fig:th}
\end{figure}
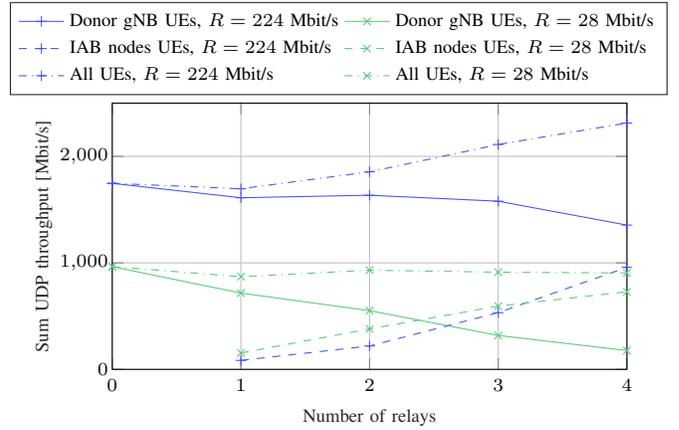

On the other hand, from Fig.~\ref{fig:lat} we notice that the average latency of the users attached to \gls{iab} nodes increases with respect to the configuration without relays, especially when  just one or two wireless relays are deployed. 
This is mainly due to the buffering that occurs in the backhaul. In an \gls{iab} context, indeed, the backhaul and access resources are shared, thus the \gls{iab} nodes and the \glspl{ue} attached to the donor contend for the same resources. With an \gls{rr} scheduler, a similar number of transmission opportunities is allocated to the \gls{iab} nodes and to the \glspl{ue}, but the relays generally have more data to transmit than each single \glspl{ue}. Consequently, the buffering latency at the \gls{rlc} layer of the relays increases.
Nonetheless, for the congested scenario (i.e.,  $R=224$ Mbit/s), the overall average latency when more than three relays are deployed (i.e., 287 and 250 ms for three and four relays, respectively) is equivalent or lower than that in the configuration with the donor \gls{gnb}~only (i.e., 292 ms), as shown in Fig.~\ref{fig:lat}.

The above discussion exemplifies how an IAB architecture  introduces both opportunities and challenges.
From one side, the deployment of  wireless relays is a viable approach to increase the coverage of cell-edge users, i.e., the most resource-constrained network entities, thereby promoting fairness in the whole network. Moreover, the presence of the wireless backhaul nodes has the potential to reduce the communication latency in case of congested networks.
From the other side, the IAB nodes may degrade the throughput and latency performance of some \glspl{ue}, i.e., those with the best channel quality, whose traffic would have been successfully handled even in traditional wired backhaul scenarios.
It becomes therefore fundamental to  determine  the optimal number of  wireless backhaul nodes to be deployed and to design efficient scheduling policies, according to the context and considering the constraints imposed by the available network and economic resources.
This research challenge will be part of our future~work.

\section{Conclusions and Future Work}
\label{sec:concl}
The integration between access and backhaul will likely play a key role in the next generation of
wireless networks operating in the  mmWave band.
In this paper, after reviewing prior work on self-backhauling and
the potential of this solution for 3GPP NR deployments, we presented the first implementation of IAB for the ns-3 mmWave module\footnote{The code can be found at \url{https://github.com/signetlabdei/ns3-mmwave-iab}.}.
The simulator, which features the 3GPP mmWave channel model and a complete characterization
of the TCP/IP protocol stack, now also implements the wireless
relaying operations on both the data and the control planes, thereby 
accurately modeling the operations of an IAB network. We believe that this tool can be used by researchers to understand the main limitations and the performance gains that \gls{iab} networks can provide, and to evaluate new integrated scheduling algorithms and multi-hop routing strategies with a realistic, end-to-end protocol stack.

We have also provided the first preliminary end-to-end performance evaluation, in terms of experienced data rate and communication latency, of mmWave nodes in an \gls{iab} scenario.
We showed that the IAB architecture may represent a viable solution to efficiently relay the traffic of cell-edge users in very congested networks.

This work opens up some particularly interesting research
directions.
More specifically, we plan to investigate how to design advanced backhaul path selection policies as well as to determine  the best
degree of migration from a fully-wired backhaul deployment
to a wireless backhaul solution when considering both economic and performance trade-offs. Moreover, we will further extend the ns-3 mmWave module with additional \gls{iab} features, in order to address mobility scenarios, and keep track of the 3GPP specifications on this~topic.

\bibliographystyle{IEEEtran}
\bibliography{bibl-abbr.bib}

\end{document}